\input ./psfig.tex
\documentclass[]{mn2e}
\def\etal{\it et al.~\rm}

\def\arcm{\hbox{$^\prime$}}
\def\arcs{\arcm\hskip -0.1em\arcm}
\def\sol{\mbox{$_{\odot}$}} 
\def\Msol{\hbox{$\thinspace M_{\odot}$}}
\def\erg{{\rm\thinspace erg}}
\def\cm{{\rm\thinspace cm}}
\def\km{{\rm\thinspace km}}
\def\s{{\rm\thinspace s}}
\def\ergps{\hbox{$\erg\s^{-1}\,$}}
\def\ecs{\hbox{$\erg\cm^{-2}\s^{-1}\,$}}
\def\kmps{\hbox{$\km\s^{-1}\,$}}
\def\h50{\hbox{h$_{50}$}}
\def\m12{\hbox{$\Delta$m$_{12}$}}
\def\lesssim{\mathrel{\hbox{\rlap{\hbox{\lower4pt\hbox{$\sim$}}}\hbox{$<$}}}}
\def\gtrsim{\mathrel{\hbox{\rlap{\hbox{\lower4pt\hbox{$\sim$}}}\hbox{$>$}}}}

\begin{document}
\title[Fossil groups of galaxies]
{The nature and space density of fossil groups of galaxies}

\author[Jones, L.R. et al.]
{L. R. Jones$^{1,6}$, T. J. Ponman$^{1}$, A. Horton$^{1}$, A. Babul$^{2}$\thanks{CITA Senior Fellow},
H. Ebeling$^{3}$ \and and  D. J. Burke$^{4}$ \\
$^{1}$School of Physics \& Astronomy, University of Birmingham, 
Birmingham B15 2TT, UK.\\
$^{2}$Department of Physics \& Astronomy, University of Victoria, Victoria, BC, V8P 1A1,
Canada\\
$^{3}$Institute for Astronomy, 2680 Woodlawn Drive, Honolulu, Hawaii 96822, USA.\\
$^{4}$Harvard-Smithsonian Center for Astrophysics, 60 Garden Street,
Cambridge, MA 02138, USA.\\
$^6$Email: lrj@star.sr.bham.ac.uk\\
}

\maketitle

\begin{abstract}
We describe the properties of a sample of galaxy groups with very
unusual distributions of galaxy luminosities.  The most extreme
example has an X-ray luminosity similar to that of the Virgo cluster
but has a very low richness, with only one galaxy brighter than L*,
compared with six in Virgo. That one galaxy, however, is optically more
luminous than any galaxy in Virgo and has an optical luminosity
as bright as many of the central cD galaxies in rich Abell clusters.

The characteristic feature of the fossil groups we study is that most
of the light arises from one dominant, central galaxy.  We define a
fossil system and, based on this definition, construct a small X-ray
selected, flux-limited sample of fossil groups with well known
selection criteria.  We confirm that these systems are indeed groups
of galaxies, but dominated by one central luminous giant elliptical
galaxy and with few, or no, L* galaxies.

We find that fossil systems represent 8\%-20\% of all systems of the
same X-ray luminosity.  Fossil groups are at least as numerous as all
poor and rich clusters combined, and are thus a possible site for the
formation of luminous central cluster galaxies before infall into
clusters occurs. The fossil systems in our sample have significantly
higher X-ray luminosities than normal groups of similar total optical
luminosities (or similar X-ray temperature, where the latter can be
measured). These enhanced X-ray luminosities may be due to relatively
cool gas in the innermost regions
or due to a low central gas entropy.

We interpret fossil groups as old, undisturbed systems which have
avoided infall into clusters, but where galaxy merging of most of the
L* galaxies has occurred. An early formation epoch, before that of
most groups, could explain low central gas entropies and high X-ray
luminosities.

\end{abstract}


\section{INTRODUCTION}

There exist concentrations of dark matter which have the gravitating
mass of groups of galaxies, and which contain X-ray emitting hot gas
with the extent and other properties expected for groups, but in which
the optical light is completely dominated by a single luminous, giant
elliptical galaxy. If systems of this type are sufficiently numerous
and massive, they may be of considerable importance as the location
of the formation of a significant fraction of all giant luminous
elliptical galaxies.  Initial indications suggest that the total mass
density of fossil groups in the Universe is similar to that of rich
clusters of galaxies (Vikhlinin \etal 1999).

Our interpretation of fossil groups is that they are the end result of
galaxy merging within a normal group, leaving behind the X-ray halo
(Ponman \etal 1994, Jones, Ponman \& Forbes 2000b; hence the term
`fossil' groups).  An alternative scenario is that they may have
formed with a very unusual galaxy luminosity distribution (Mulchaey \&
Zabludoff 1999), although we will show that this is a less likely
origin. If the merger interpretation is basically correct, 
the timescales for dynamical friction on
L$^*$ galaxies suggest that fossil groups are old, undisturbed systems
that have seen little infall of L$^*$ galaxies since their initial collapse.
Fossil groups are thus an important laboratory for studying the
formation and evolution of galaxies and the intra-group medium in an
isolated system.  They may also be the evolutionary link between
compact galaxy groups and giant ellipticals (possibly via a ULIRG
phase; Borne \etal 2000).  In addition, these systems act as a
reminder that not all mass concentrations can be found by optical
means; the total mass and spatial extent of fossil groups has only
been revealed via X-ray observations.

It is thus important to measure the space density of fossil systems.
Several candidate systems have been reported, but the criteria used to
define this class of objects vary and the two existing estimates of
their space density disagree by a factor of $\sim$4 (Vikhlinin \etal
1999, Romer \etal 2000).  Vikhlinin \etal (1999) found four potential
fossil systems which they considered to be isolated galaxies, calling
them X-ray over-luminous elliptical galaxies (OLEGs). Ponman \etal
(1994) and Jones \etal (2000b) have studied one of these systems in
detail and found that it is in reality a group, with a halo of dwarf
galaxies, and a deficit of L$^*$ galaxies.  Mulchaey \& Zabludoff
(1999) found a similar distribution of galaxy luminosities around an
optically selected giant elliptical galaxy with an over-luminous X-ray
halo. Matsushita \etal (1998) and Matsushita (2001) may have found 
similar systems which they call X-ray extended early type galaxies.

In this paper we adopt well-defined selection criteria for fossil groups
and measure their space density as a function of X-ray luminosity. We
also compare their X-ray luminosities and temperatures with those of
other groups and clusters.  We assume H$_0$=50\h50 \kmps Mpc$^{-1}$
and q$_0$=0.5 throughout.

\section{Definition of a fossil system}

We define a fossil system in observational terms as a spatially
extended X-ray source with an X-ray luminosity from diffuse, hot gas
of $L_{X,bol}\geq 10^{42}$ \h50$^{-2}$ \ergps.  The optical
counterpart is a bound system of galaxies with \m12$\geq$2.0 mag,
where \m12 is the absolute total magnitude gap in R between the
brightest and second brightest galaxies in the system within half the
(projected) virial radius (r$_{vir}$).  No upper limit is placed on
the X-ray luminosity or temperature. 

The rationale behind this choice of definition is as follows. The
lower limit in X-ray luminosity applied by us excludes normal
galaxies. O'Sullivan, Forbes \& Ponman (2001) find very few elliptical
galaxies with higher X-ray luminosities that are not at the centres of
groups, and thus may have a contribution to their measured $L_X$ from
the hot gas of the group (see also  Matsushita 2001). The limit in \m12 ensures that the
brightest galaxy dominates the system. We use half the virial radius
because this corresponds approximately to the radius within which
orbital decay by dynamical friction (Binney \& Tremaine 1987) predicts
that M$^*$ galaxies (with mass/light ratio = 10\Msol/L\sol) will fall
to the centre of the system within a Hubble time. The precise value
of the threshold in \m12 is somewhat arbitrary, but we show in Section 
5.3 that the probability of
obtaining \m12$>$2 by chance from a typical Schechter function is very
small. Observationally, it is also unusual to find groups or clusters of galaxies
with \m12$>$2.  For example, of the 24 optically selected poor
clusters of Price \etal (1991), the highest value of \m12 is
1.3. Similarly, of the 20 MKW (Morgan \etal 1975) and AWM (Albert
\etal 1977) poor clusters of Beers \etal (1995), the highest value of
\m12 is 1.6. The MKW/AWM clusters were selected to contain a dominant
giant elliptical galaxy, and thus if anything should have larger
values of \m12 than randomly selected clusters. The MKW/AWM clusters
also cover the range of X-ray luminosity of the fossil systems studied
here.

\section{Initial sample selection}

To search for fossil groups we used our existing X-ray selected sample
of spatially extended, serendipitous X-ray sources compiled during the
WARPS project (Wide Angle $ROSAT$ Pointed Survey; Scharf \etal 1997,
Jones \etal 1998, Perlman \etal 2002). This flux limited survey
contains $\approx$150 extended X-ray sources, from nearby individual
galaxies to clusters of galaxies at high redshifts (Ebeling \etal
2000, 2001), all detected in the central regions of $ROSAT$ PSPC
(Position Sensitive Proportional Counter) pointings at high Galactic
latitudes ($|$b$|>20^{\circ}$), in the 0.5-2 keV energy band. Because
of the crucial need to be able to identify sources as extended, only
the inner region of 15 arcmin radius of the detector was used. The
variation in the full width at half maximum (FWHM) of the instrumental
point spread function (PSF) was thus restricted from values of 25\arcs~
on-axis to a maximum of 45\arcs~ at the largest radius.  Most of the
targets of these pointings were either Galactic objects or AGN.  In
order to maximise the available sky area, we have used all 86 $ROSAT$
fields in phase one of WARPS (as described in Jones \etal 1998 and
Perlman \etal 2002), including five fields which had clusters of
galaxies as targets, combined with the much larger number of 303
$ROSAT$ fields from phase two of WARPS. The total area of sky surveyed
was 73 deg$^{2}$.

Source detection and characterisation was based on the Voronoi
Tesselation and Percolation technique (Ebeling \& Wiedenmann 1993) and
is described by Scharf \etal (1997). 
The number of counts in each detection varies between 60-160 (0.5-2 keV),
except for RXJ1416.4+2315 which was detected with $\approx$800 counts.
To extrapolate from the detected
count rate to a total count rate, allowing for the undetected flux
below the limiting X-ray isophote, we assumed a King profile with
$\beta$=2/3, a core radius estimated from the PSPC data, and
extrapolated to infinite radius (extrapolating to the virial radius
would give count rates only $\approx$10\% lower).  The resulting mean
increase in the count rate for the confirmed fossil systems described
below was 53\%.  If $\beta\approx$0.5, as found for many groups
(Helsdon \& Ponman 2000), then the luminosities would be slightly
higher (a correction factor $\approx$2, rather than 1.5).  The
conversion from total PSPC count rate to flux in the 0.5-2 keV band
included a small correction for Galactic absorption in the direction
of each source and used a temperature estimated from an empirical
$L_X-T$ relation (Perlman \etal 2002). The dependence of the flux on
the precise value of this temperature was small ($\approx$6\%; Jones
\etal 1998).  We employ a limit of 5.5$\times 10^{-14}$ \ecs (0.5-2 keV) in
total flux, slightly lower than in the full WARPS catalogue, to
maximise the volume surveyed for the relatively rare fossil systems of
galaxies.

To derive a list of candidate fossil systems, all X-ray sources
flagged as extended or possibly extended (see Scharf \etal 1997) were
investigated using a combination of methods.  This work started whilst
spectroscopic identification of extended sources in the full WARPS
catalogue
was not yet complete, and so an initial method was devised to help
identify fossil candidates, for which spectroscopic identifications
have subsequently been obtained.  The virtually complete spectroscopic
identification of the full WARPS sample, using methods described in
Jones \etal (1998) and Perlman \etal (2002), has since been used to
confirm and update the list of candidate fossil systems.

The initial method combined visual examination of X-ray/optical
overlays of all the sources, guided additionally by HRI data of higher
resolution where available, optical spectroscopic identifications,
R-band CCD images, identifications in NED, and the ratio of X-ray to
optical flux.  This search used POSS I \& II Digitised Sky Survey
data, plus deeper R-band CCD images for the majority of the extended
sources.  The ratio of X-ray to optical flux helped to distinguish
cases where close point X-ray sources (typically AGN) were blended
together, masquerading as an extended source.  Particular attention
was paid to sources with log($f_x/f_{opt})<-0.8$, where $f_x$ is the
0.5-2 keV X-ray flux (in \ecs) and

\begin{equation}
  f_{opt}=2.2$x$10^6$ x $10^{(-0.4R - 11.76)}  \ecs  
\end{equation}
(Allen 1973), where the R magnitude is that of the counterpart of the
X-ray source, or of the brightest galaxy in the system for clusters
and groups of galaxies. Most AGN and normal clusters had
log($f_x/f_{opt})>-0.8$, whereas fossil systems had lower values of
log($f_x/f_{opt}$).  Where X-ray emission from both a hot inter-group
medium (IGM) and an AGN in the group was observed, an estimate of the
IGM flux alone was made.  Initially APM POSS E magnitude estimates
(Irwin, Maddox \& McMahon 1994), accurate to $\approx$0.25 mag, were
used instead of true R-band magnitudes.  For very bright galaxies, the
APM magnitudes were sometimes underestimated (ie brighter than in
reality), but since this decreased $f_x/f_{opt}$ and increased \m12,
it only increased the number of candidates.

We emphasise that the selection of the final list of 18 candidate
fossil systems selected for further scrutiny was based on a {\it
variety}\/ of indicators: optical spectroscopy and classification of
the system in the full WARPS survey, but also the visual appearance
(is the system dominated by a single galaxy?) and the $f_x/f_{opt}$
value. We have thus been conservative in our selection of candidate
systems, and it is unlikely that we have missed a significant number,
particularly at low redshifts (the completeness of our sample is
discussed further in Section 4.2).

\subsection{Further observations and analysis of the candidate systems}

Although one to three redshifts of member galaxies were obtained as
part of the WARPS survey, more redshifts were required for many
systems where further galaxies within \m12=2.0 mag of the brightest
galaxy were located within projected radii $<0.5$r$_{vir}$.  It was
necessary to know if these galaxies were group members or not, in
order to determine if the systems met our definition of a fossil
system.  Redshifts were obtained with the RC spectrograph at the Kitt
Peak 4m telescope, using a resolution of 7$\AA$. Data reduction
followed standard procedures using $IRAF$. Further details of the
imaging and spectroscopic observations will be given elsewhere.

CCD imaging of the most promising candidates, to R=24, was obtained in
service time with the 2.5m Isaac Newton Telescope (INT) wide-field
camera to supplement existing, shallower images obtained as part of
the WARPS survey.  Data reduction was again performed in the standard
way, including a correction for the small non-linearity of the CCDs.
Unfortunately the conditions were not photometric, and so further
R-band imaging was obtained, in photometric conditions, to calibrate
the original images.  For two systems (J1416.4+2315 and J1552.2+2013)
this further imaging was obtained with the 8k mosaic camera at the
University of Hawaii 2.2m telescope.  For the other systems the
calibration images were obtained in additional INT wide field camera
service time.  All photometric calibration was based on observations
of 5-12 standard stars per night selected from Landolt (1992).  The
resultant photometric accuracy for all the systems is $\approx$0.05
mag.

Accurate \m12 values for the candidate fossil systems were measured
from the R-band CCD images. For each galaxy, an aperture containing
$>90\%$ of the total light, as judged from the curve of growth, was
used to measure a pseudo-total magnitude. Absolute magnitudes of the
central galaxies were corrected for Galactic absorption and were
K-corrected.  No K-corrections were applied in deriving \m12, since
the corrections are small ($\approx$0.1 mag) at the redshifts sampled,
and in any case all of the brightest galaxies, and most of the second
brightest galaxies, are of an early type and would have identical
K-corrections.

The virial radii were estimated using the relationship between virial
radius and X-ray temperature r$_{vir}$=r$_{200}$=3.89 (T/10
keV)$^{0.5}$(1+z)$^{-1.5}$ h$_{50}^{-1}$ Mpc (Evrard \etal 1996). For
most systems, there were too few $ROSAT$ PSPC counts to reliably
measure a temperature, and for these a temperature estimate was made
based on the luminosity-temperature (L$_X$-T) relations of White \etal
(1997) and Helsdon \& Ponman (2000).  The virial radii estimated in
this way varied between 0.94-1.45 h$_{50}^{-1}$ Mpc.  We show below
that at least some fossil systems lie on the low-temperature side of
the the L$_X$-T relation, so that our estimates of both temperature
and virial radius for systems without reliable temperature
measurements may be overestimates.  Reducing the virial radii, and
thus the sky area where \m12 is measured, can only increase \m12 and
reinforce the fossil nature of the systems.  For one system
(RXJ1416.4+2315) there were sufficient $ROSAT$ PSPC counts to measure
a temperature, as described by Fairley \etal (2000).  Absorption was
fixed at the Galactic value, and abundances fixed at 0.3 x solar. The
best fit MEKAL (Mewe \etal 1986) temperature was
1.53$^{+0.48}_{-0.19}$ keV (at 90\% confidence).

Radial profiles of the X-ray surface brightness as observed with the
$ROSAT$ PSPC (and HRI where available) were compared to the
appropriate PSFs, and only genuinely extended X-ray sources were
accepted as fossil systems.

After combining the new redshift information, photometry, and virial
radius estimates, 5 new fossil systems were identified which meet the
definition given in Section 2 (diffuse, hot gas of luminosity
$L_{X,bol}\geq10^{42}$ \h50$^{-2}$ \ergps, and \m12$\geq$2.0 mag
within 0.5r$_{vir}$). Twelve candidates were ruled out, and one
candidate remains, awaiting further spectroscopy.

\section{Results}

\subsection{The fossil systems}

The five new fossil systems are shown in Fig 1. Their basic
properties, together with those of the one remaining candidate system,
are given in Table 1. Here we provide brief descriptions; the results
of detailed observations will be reported elsewhere.  The system
RXJ1340.6+4018, described by Jones \etal (2000b), meets the fossil
group definition used here, but it is not included in the present 
statistical sample because the
exposure time of the $ROSAT$ PSPC observation in which it was
discovered was below the threshold used in the WARPS survey.

\begin{figure}
\psfig{figure=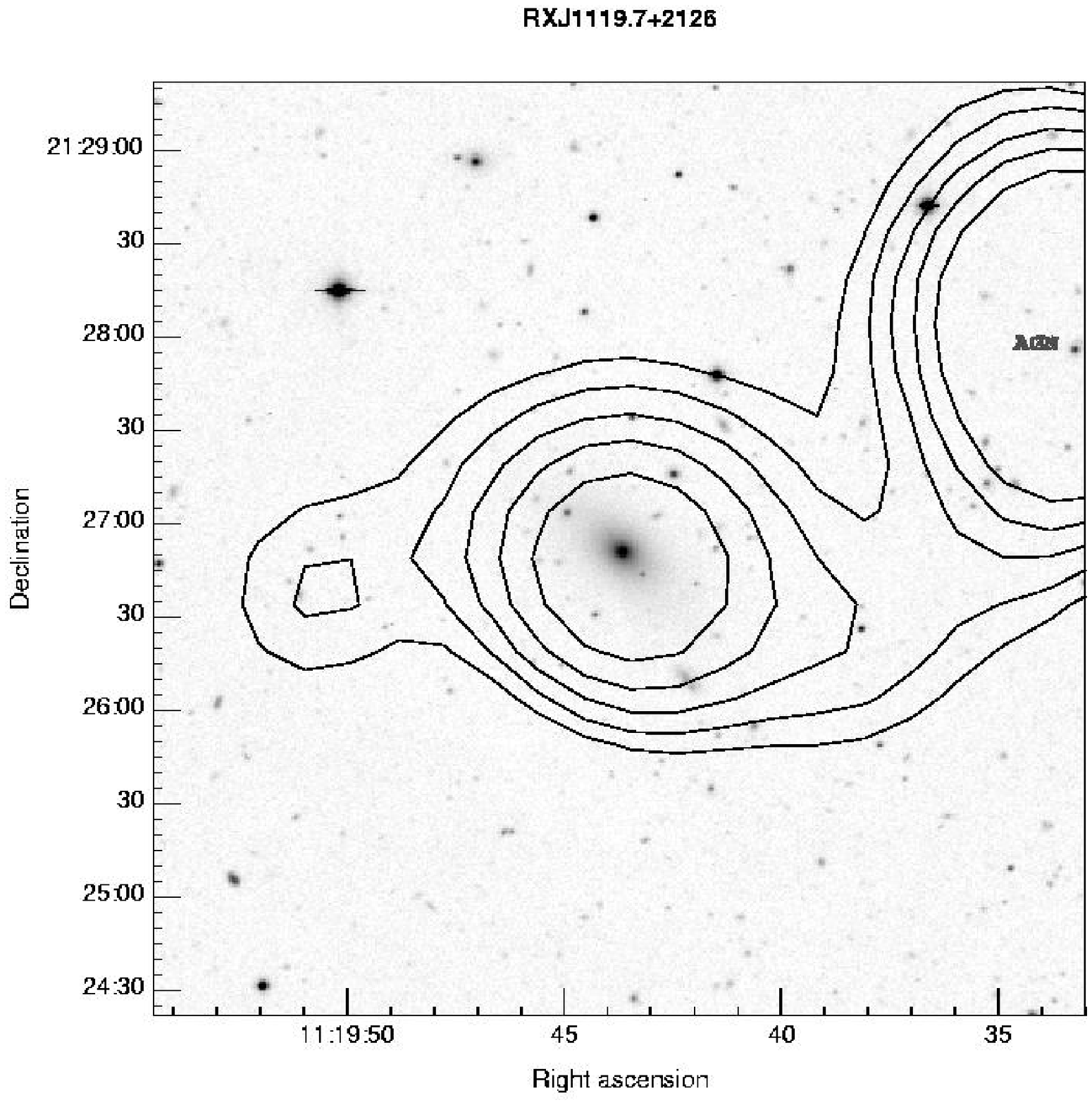,height=8.0cm,angle=0}
\caption{$R$-band CCD images and overlaid X-ray contours from the $ROSAT$ 
PSPC (0.5-2 keV) of the fossil systems.  The greyscale intensity
scaling is logarithmic.  The X-ray images have been adaptively
smoothed and contoured at intervals of a factor of 1.4 in surface
brightness, except for RXJ1119.7+2126, where the interval is a factor
of 1.2.  }
\end{figure}

\vspace{1cm}
\begin{figure}
\psfig{figure=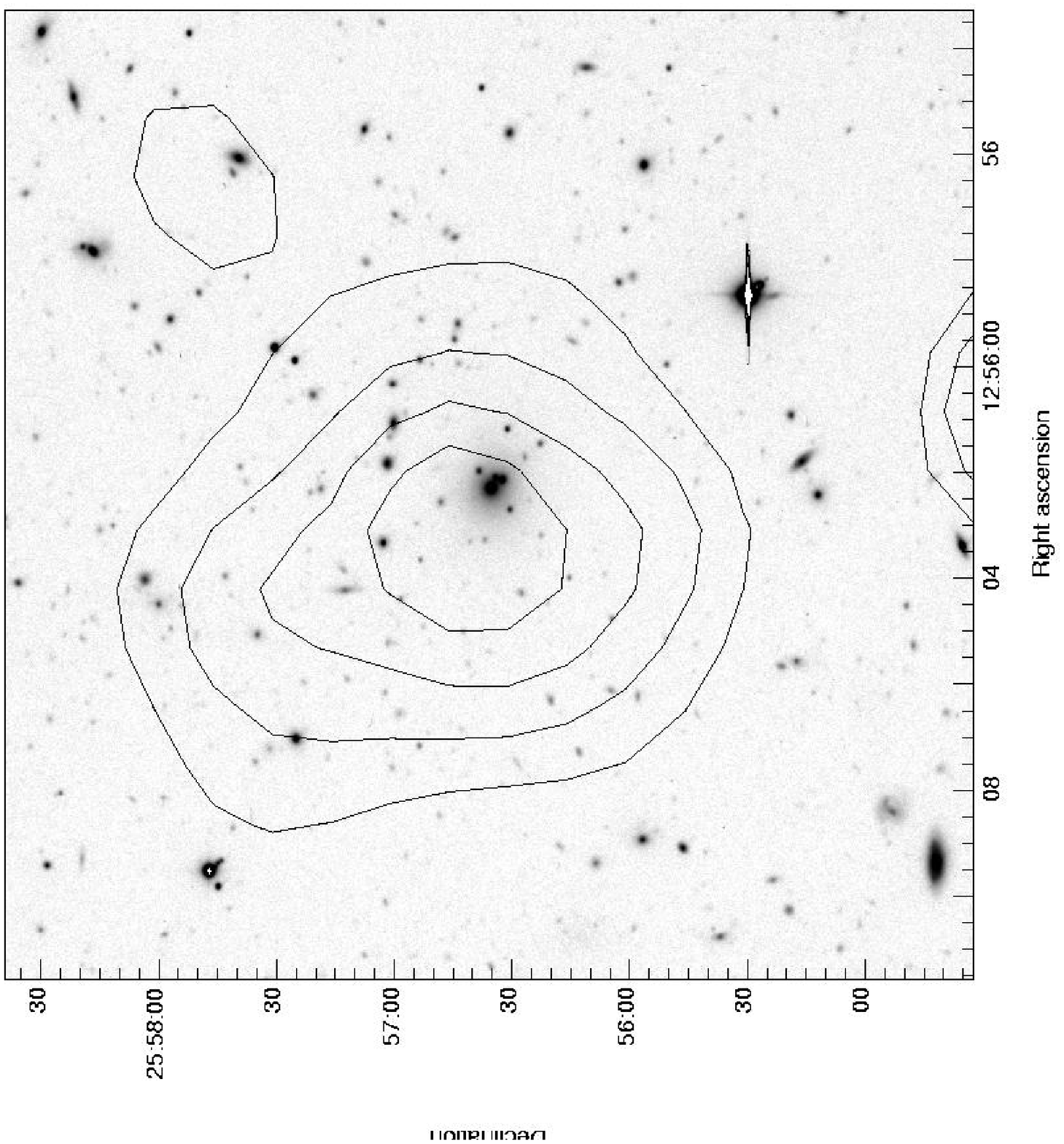,height=7.0cm,angle=-90}
\end{figure}

\begin{figure}
\psfig{figure=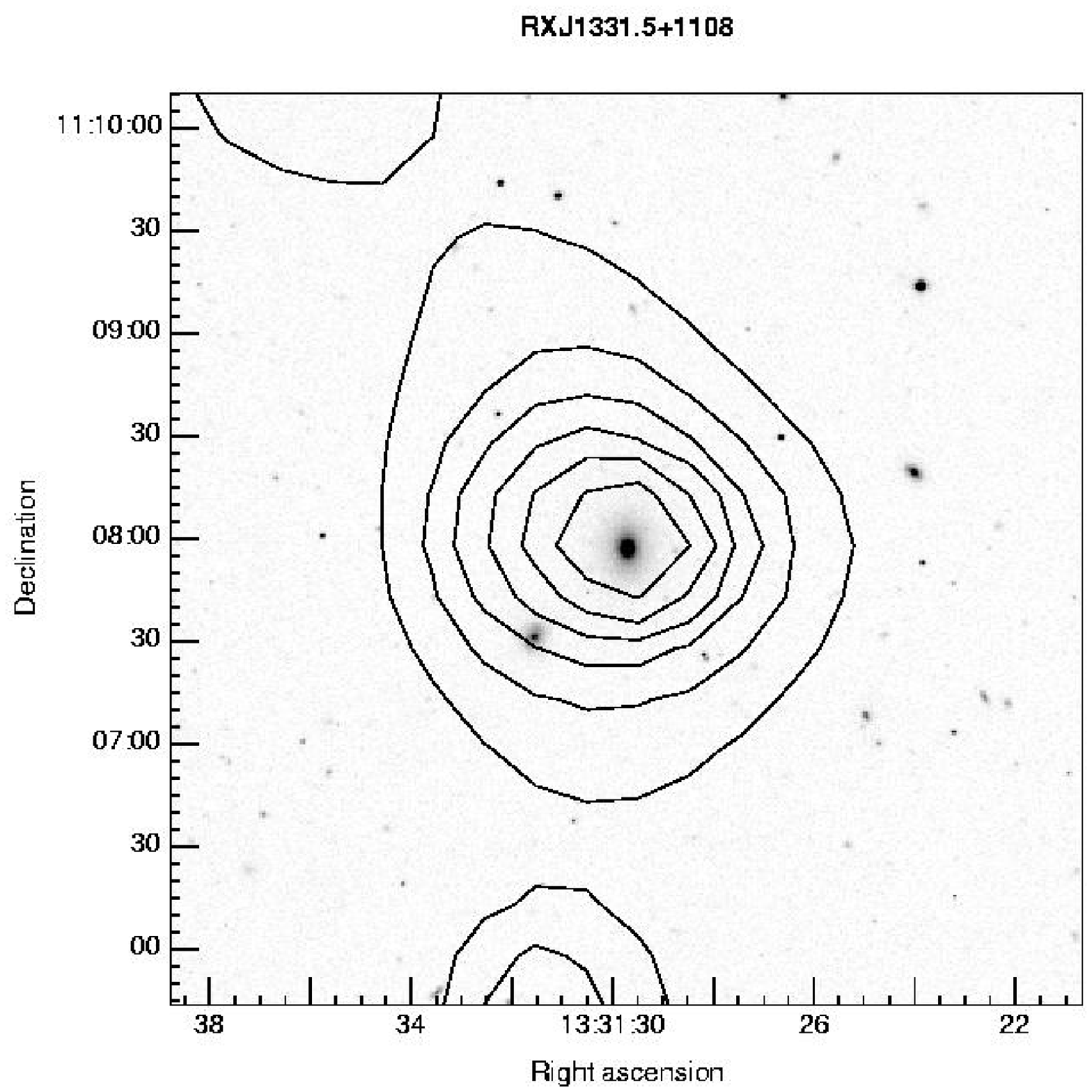,height=8.5cm,angle=0}
\end{figure}

\begin{figure}
\psfig{figure=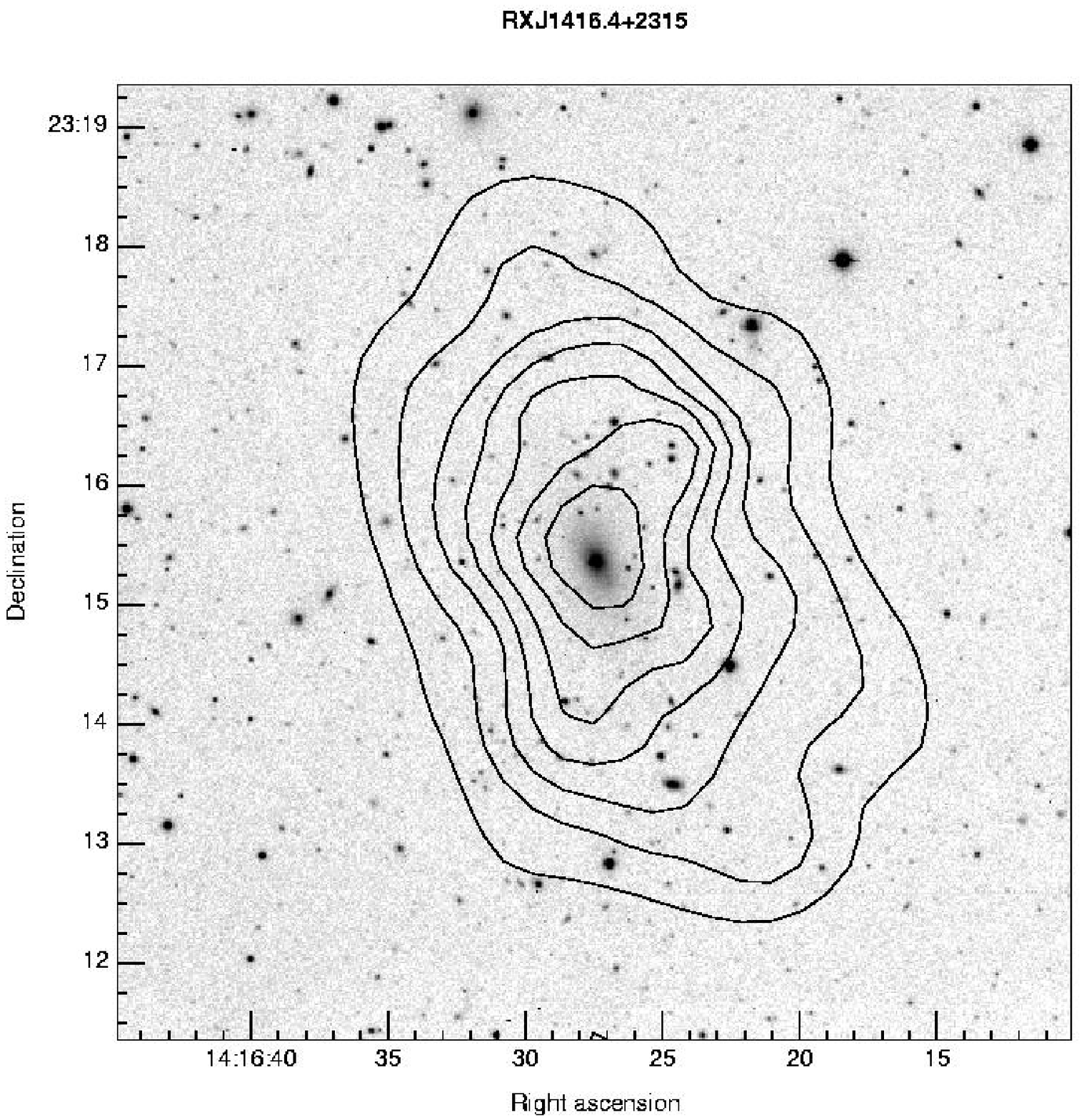,height=8.0cm,angle=0}
\end{figure}

\begin{figure}
\psfig{figure=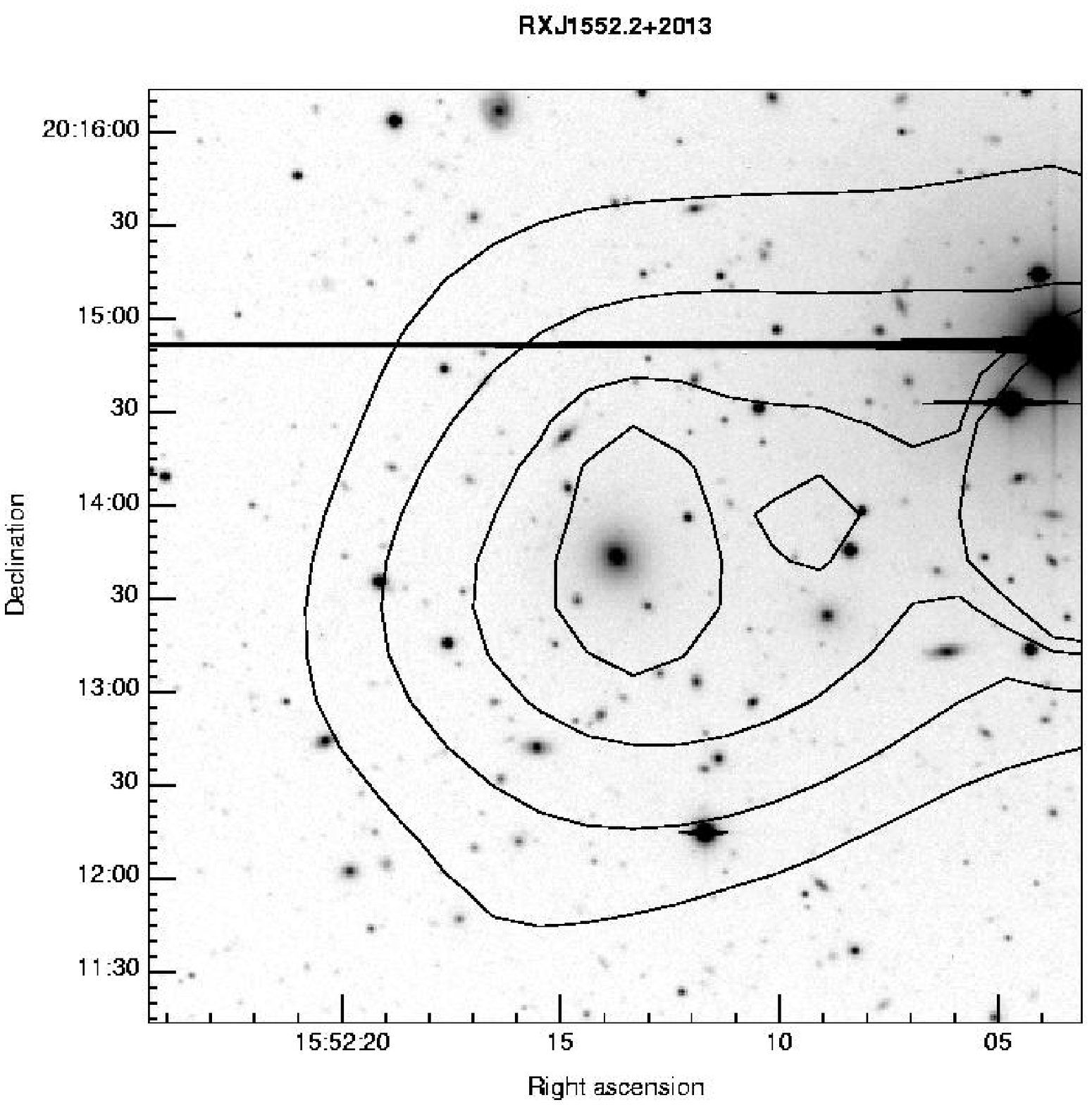,height=8.0cm,angle=0}
\end{figure}

\noindent{\bf RXJ1119.7+2126.}
This is the lowest-luminosity fossil group in the sample. Optically,
the group is completely dominated by a giant elliptical galaxy at a
position coincident with the peak of the X-ray emission. This galaxy
has M$_R$=-22.8 + 5logh$_{50}$ and is visible to a semi-major axis
distance of 70 h$_{50}^{-1}$ kpc in our deep CCD image. At least two
nearby galaxies, 3.3-3.4 mag fainter, are spectroscopically confirmed
group members, and there is an excess of faint galaxies visible within
the X-ray contours.  The PSPC data within a radius of 1 arcmin of the
peak emission are clearly extended. The quoted X-ray luminosity arises
only from the source centred on the bright elliptical galaxy, and
excludes the nearby point X-ray sources visible in Fig 1 (one of which
is labelled and is an AGN at z=0.282; Mason \etal 2000).  We find that
\m12 is definitely greater than two and lies in the range 2.5-3.3,
depending on the group membership of one particular galaxy.  This
source is also identified as a cluster/group of galaxies in the RIXOS
(Mason \etal 2000) and Vikhlinin \etal (1998) surveys.  Mason \etal
(2000) and Tripp \etal (1998) confirm the redshift; Vikhlinin \etal
had no spectroscopic redshift.

\noindent{\bf RXJ1256.0+2556.}
The emission seen in the $ROSAT$ PSPC data extends to a radius of 2
arcmin and is clearly extended.  Optically this source is a rich group
of galaxies, dominated by a luminous (M$_R$=-24.8 + 5logh$_{50}$)
elliptical galaxy, with fainter galaxies observed in projection
against it.  The offset of $\approx$15\arcs of the peak of the X-ray
emission compared to the brightest galaxy is not significant given the
positional uncertainty of $\approx$10\arcs of the ROSAT attitude
solution.  The classification as a fossil system is somewhat uncertain
because \m12 may be either 1.6 or $>$2.0, depending on whether a group
galaxy at a radius within, but close to, 0.5r$_{vir}$ is included.
The value of the virial radius is uncertain since the X-ray
temperature is unknown. For the purposes of this paper we include this
system, but note that it contains at least 1-2 L$^*$ galaxies at a
radius $\approx$0.5r$_{vir}$.  All nearby point-like X-ray sources
visible in Fig 1 are excluded from the quoted luminosity.  This X-ray
source is also identified as a cluster/group of galaxies in the
Vikhlinin \etal (1998) survey, but without a spectroscopic redshift.

\noindent{\bf RXJ1331.5+1108.}
The X-ray luminosity of diffuse, hot gas in this source is slightly
uncertain because of the possibility of point source contamination,
but it is almost certainly above 10$^{42}$ \ergps.
Optically it is a fossil group: we have 3 spectroscopically
confirmed group members and \m12=2.0 mag.  The source is only
marginally extended in the PSPC. However, two HRI observations, both
with sufficient sensitivity to detect a point X-ray source with the
PSPC flux at the position of the central galaxy, fail to do so. The
required variability of more than a factor of 2 on a timescale of 1
month is not unlikely for an AGN. However, the summed HRI observations
do reveal faint diffuse emission (possibly including a
very faint point source) over a $\sim$1 arcmin$^2$ region adjacent to
the central galaxy. This faint HRI emission has a flux $\geq$50\% of
the total PSPC flux.

The central galaxy, at the X-ray peak, has narrow H$\alpha$ and [SII]
emission lines.  An unresolved (r$<$2\arcs) 12.6$\pm$0.6 mJy radio
source at 1.4 GHz is coincident with the central galaxy (from the
FIRST survey; White \etal 1997). This flux corresponds to a a power of
$4.1\times 10^{30}$ erg s$^{-1}$ Hz$^{-1}$ at 1.4 GHz, similar to that of
the first fossil system found by Ponman \etal (1994) and Jones \etal
(2000b), and comparable to values found for radio-loud cD galaxies in
cluster cores.  These cluster radio sources are often associated with
cooling flows, and the emission line spectrum may also be associated
with a cooling flow. In the RIXOS survey, based only on PSPC data and
one galaxy redshift, Mason \etal (2000) classify this source as a
Sy2/liner. With the additional data discussed here, we classify the
source as a probable combination of fossil group and AGN, with the
X-ray luminosity of each component remaining uncertain.


\noindent{\bf RXJ1416.4+2315.}
This is the most X-ray luminous source in the sample. In a deep
optical image it would be classified as a galaxy group or poor cluster
centred on an extremely dominant, luminous, giant elliptical
galaxy. This remarkable galaxy has M$_R$=-25 + 5logh$_{50}$ and is
visible to a semi-major axis length of 160 h$_{50}^{-1}$ kpc.  The
very extended X-ray emission is detected to a semi-major axis length
of 3.5 arcmin (650 h$_{50}^{-1}$ kpc) in the PSPC and is elongated in
a direction similar to that of the central elliptical galaxy.

A radio source of flux 3.4$\pm$0.2 mJy at 1.4 GHz is coincident with
the central galaxy. The radio source is extended on a scale of 4
arcsec (White \etal 1997).

This source appears in the bright SHARC cluster survey of Romer \etal
(2000).  The redshift is confirmed, but it is not identified as a
fossil system by Romer \etal.  There is a small ($<$10\%) amount of
contamination from a background QSO (HS 1414+2330 at z=1.54, Hagen
\etal 1999) in the PSPC image, which we have not removed.  The target
of this $ROSAT$ field was a candidate cluster of galaxies, but at a
high redshift (z$>$0.3), and unrelated to the fossil system.

\noindent{\bf RXJ1552.2+2013.}
The second most X-ray luminous system in the sample, this clearly
extended X-ray source has the optical appearance of a galaxy group,
again dominated by a giant luminous elliptical galaxy at the X-ray
peak.

The bright X-ray point source 2.5 arcmin west of RXJ1552.2+2013 is a
background QSO at z=0.25. The X-ray flux from this QSO does not
contribute to the luminosity given in Table 1, but a small amount of
contamination ($\sim10$\%) from a probable AGN in the group
approximately 1 arcmin W of the central galaxy is included in the
quoted luminosity. No point-like X-ray sources are detected within the
PSPC contours in an HRI exposure with sufficient sensitivity to detect
any point sources with $\approx$15\% of the total PSPC flux.

This source is identified as a group/cluster of galaxies in the survey
of Vikhlinin \etal (1998), who confirm the redshift, and as an
extended X-ray source, but below the count rate limit, in the survey
of Romer \etal (2000).

\noindent{The final system described here is the remaining candidate.}\\
\noindent{\bf RXJ0116.6-0329.}
Although this is a clearly extended PSPC source identified with a
group of galaxies with two spectroscopically confirmed redshifts, the
value of \m12 is very uncertain. There are 3-4 bright galaxies at a
radius of approximately 0.25r$_{vir}$ with unknown redshifts.  If any
one of these is a group member, then \m12$<$2.0, and this would not be
a fossil system as defined here. Thus we exclude this source from
further discussion.
 
\begin{table*}
\begin{minipage}{175mm}
\caption {Basic properties of the fossil galaxy groups}

 \begin{tabular}{lllllllllll} \hline
Name & RA (J2000)& Dec & z & n$_z$ & \m12 & f$_{X}$(0.5-2 keV)  &\multicolumn{2}{|c|}
{L$_{X}$ (10$^{42}$ \h50$^{-2}$}  & T$_X$&  M$^{BCG}_{R}$ \\
     &           &     &   &       & (mag)& (10$^{-13}$ & 
\multicolumn{2}{|c|}{\ergps)}  &
(keV)& (+5log(h$_{50}$))\\
                  &            &           &       &(a)&   (b)   & \ecs)&(0.5-2 keV)& (bol)    &                          &      \\
                  &            &           &       &   &         &      &          &           &                          &      \\
RXJ1119.7+2126    & 11 19 43.7 & +21 26 50 & 0.061 & 3 & 3.3$^c$ & 0.57 &  1.0     &   1.7     & -                        & -22.8\\
RXJ1256.0+2556    & 12 56 03.4 & +25 56 48 & 0.232 & 3 & 2.5$^c$ & 1.06 & 26.      &  61.      & -                        & -24.8\\
RXJ1331.5+1108    & 13 31 30.2 & +11 08 04 & 0.081 & 3 & 2.0     & 1.07 &  3.2$^d$ &   5.9$^d$ & -                        & -23.6\\
RXJ1416.4+2315    & 14 16 26.9 & +23 15 32 & 0.137 & 6 & 2.4     & 12.5 &103.      & 220.      & 1.53$^{+0.48 e}_{-0.19}$ & -25.0\\
RXJ1552.2+2013    & 15 52 12.5 & +20 13 32 & 0.135 & 4 & 2.3     &  3.32& 27.      &  63.      & -                        & -24.7\\
& & & & & & & & & & \\
RXJ0116.6-0329$^f$& 01 16 40.2 & -03 29 57 & 0.081 & 2 & ?       &  2.90&  8.3     &   17.     & -                        & -23.8\\ \hline
\end{tabular}

\small

(a) Number of spectroscopically confirmed member galaxies\\
(b) Magnitude gap in R between the brightest and second brightest
galaxy members within a projected radius of 0.5r$_{virial}$.\\
(c) \m12 uncertain. See text for details.\\  
(d) May be an overestimate due to a contribution from an AGN.\\
(e) 90\% confidence limits\\
(f) Unconfirmed, unlikely to be a fossil system; \m12 unknown.\\ 
\normalsize
\end{minipage}
\end{table*}

\subsection{Selection effects and completeness}

Although our definition of a fossil system placed no upper limit on
the X-ray luminosity, and thus in principle very luminous fossil
clusters of galaxies, as well as groups, were detectable, in practice
some selection effects were at work. In a survey with a single flux
limit, the most luminous sources are found at moderate to high
redshifts, where the search volume is largest.  However, at the
highest redshifts the depth of some of the imaging and spectroscopic
follow-up was insufficient to reject galaxies with \m12=2.0 mag as
non-cluster members.
Combined with the increase in numbers of projected foreground and
background galaxies with redshift, this meant that a reliable sample
of high-redshift fossil systems, as defined here, was impractical. We
set a conservative maximum redshift of z=0.25 within which this survey
should be reasonably complete since the spectroscopy is deep enough to
reliably identify fossil systems; indeed the highest redshift fossil
system found is at z=0.232.

The survey flux limit implies that fossil candidates with
L$_X=10^{42}$ \ergps (0.5-2 keV) will be included in the sample only
at low redshifts z$<$0.06.  Systems of L$_X>10^{43}$ \ergps will be
included at all redshifts up to the limit of z=0.25.  Relatively few
high-luminosity ($\sim10^{45}$ \ergps) systems were detected in WARPS
at z$<$0.25 (because of the relatively small volume at these redshifts),
so it is not surprising that no very luminous fossil
clusters are in the sample, if they exist at all.

To check if any  incompleteness can be found, we make a direct comparison with the two
surveys in which systems similar to our fossil groups were identified,
also based on $ROSAT$ serendipitous detections (Vikhlinin \etal 1999;
Romer \etal 2000). Unfortunately, there is little overlap.  Of the
four systems indentified by Vikhlinin \etal, only one is in a $ROSAT$
field which is also contained in the WARPS survey, but it falls just
outside our maximum radius of 15 arcmin and thus was not considered in
the survey.  Similarly, of the three fossil systems identified by
Romer \etal 2000, only one is in a $ROSAT$ field which overlaps with
this survey, but again it is outside our maximum radius of 15 arcmin.

Four of the fossil groups found here are in fields used in the
Vikhlinin \etal (1998) survey, which was the basis of the Vikhlinin
\etal (1999) overluminous elliptical galaxy (OLEG) sample, but none
are included in that OLEG sample.  The X-ray luminosities of two
fossil groups are lower than the limit applied by Vikhlinin \etal
(1999), explaining their absence.  The other two missing systems may
be explained by the different selection criteria applied by Vikhlinin
\etal (1999).  One of them (RXJ1552.2+2013) does not appear to be a
fossil group until redshifts of nearby bright galaxies reveal that
they are in the foreground, perhaps explaining its absence, and the
other (RXJ1331.5+1108) may be contaminated by an X-ray point
source, as already noted.

\subsection{The space density of fossil groups}

The space density was estimated using the $1/V_a$ statistic of Avni \&
Bahcall (1980).  The survey volume within z=0.25 was calculated as a
function of luminosity using the sky area available at any given total
flux, including the X-ray K-corrections of Jones \etal (1998).  The
sky area was calculated using simulated cluster profiles and the
distribution of PSPC exposure times and background levels in the
survey, as described in Scharf \etal (1997).  In fact three of the five
fossil groups have sufficiently high fluxes to be detectable in almost
all the survey fields, so that the details of the area calculation are
unimportant for these three.

The integrated space densities we find at different limiting
luminosities, as well as those of previous studies, are given in Table
2.  Given the small number statistics, these values are in reasonable
agreement with the Vikhlinin \etal (1999) space density 
extrapolated to lower luminosities.  The Romer \etal (2000) space
density is, however, noticeably higher than the other two estimates.

An estimate of the differential X-ray luminosity function is shown in
Fig 2.  Although the small sample size has required a large bin size
in luminosity to be used ($\sim$1 decade in L$_X$), it is clear that
fossil groups are more numerous than Hickson Compact Groups (HCGs)
and, by definition, are less numerous than all clusters and groups
at a similar L$_X$ combined.

\begin{table}
\caption {Integrated space densities of     fossil galaxy groups}

 \begin{tabular}{llll} \hline
$L_{X}$$^a$ (10$^{42}$ \h50$^{-2}$  & N$_{foss}$$^{(b)}$  & $N(>L_{X})$ &  Reference$^{(c)}$\\
\ergps, 0.5-2 keV) & &  \h50$^{3}$ Mpc$^{-3}$ &   \\

  &  & & \\
$>$1          & 5 & 4$^{+2.7}_{-1.8}\times 10^{-6}$ & this work \\
$>$10         & 3 & 2$^{+1.9}_{-1.1}\times 10^{-7}$ & this work \\
$>$20         & 4 & 2.4$^{+3.1}_{-1.2}\times 10^{-7}$ $^{(d)}$  & V99 \\
$>$10$^{(e)}$ & 4 & 4.6$^{+5.9}_{-2.3}\times 10^{-7}$ $^{(d)}$  & V99 \\
$>$10         & 3 & $\sim 2\times 10^{-6}$ & R00 \\

             \hline
\end{tabular}

\small
(a) Limiting X-ray luminosity\\
(b) Number of fossil systems\\
(c) V99: Vikhlinin \etal (1999), R00: Romer \etal (2000)\\
(d) 90\% confidence limits quoted by Vikhlinin \etal (1999)\\
(e) extrapolated from $2\times10^{43}$ \ergps to $1\times 10^{43}$
\ergps assuming the same slope as the cluster X-ray luminosity
function.\\

\normalsize
\end{table}

\begin{figure}
\psfig{figure=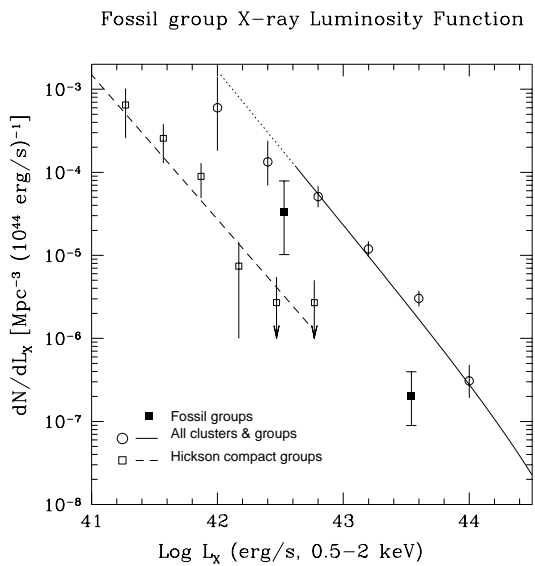,height=8.0cm,angle=0}
\caption{Differential X-ray luminosity function estimate for fossil groups 
(solid squares) compared with all clusters and groups (solid line and
open circles; Ebeling \etal 1997 \& Jones \etal 2000a), and with
Hickson compact groups (open squares and dashed line; Ponman \etal
1996). The Ponman \etal HCG data have been shifted by a fixed factor
of 0.53 in luminosity to convert from bolometric luminosities to the
0.5-2 keV band. This factor is appropriate for a mean HCG temperature
of 1 keV.  }
\end{figure}

\subsection{The L$_X$-T relation}

Only one of the systems described here has sufficient $ROSAT$ counts
to measure a temperature. This system has an excess X-ray luminosity
for its temperature, as also found for the fossil group RXJ1340.6+4018
(Jones \etal 2000b). In Fig 3 we plot both these fossil systems on the
low-redshift X-ray luminosity-temperature relation for groups and
clusters obtained from a variety of sources. Both fossil systems have
the highest X-ray luminosity of all the systems in the plot at that
temperature.  Most groups are {\it underluminous}\/ compared to an
extrapolation of the cluster $L_X-T$ relation, whereas the opposite is
true for these two fossil groups. The luminosities of all the low
temperature systems in Fig 3 have been extrapolated to the virial
radius.

It remains to be seen whether {\it all}\/ fossil systems are
overluminous.  Vikhlinin \etal (1999) find that one OLEG is consistent
with the group $L_X-T$ relation.  It is possible that there is a
selection effect in our sample in the sense that we were only able to
measure temperatures for the most luminous systems, because they gave
the highest-quality spectra.  However, the positions of these fossil
groups on the $L_X-T$ relation are clearly extreme compared to a sizeable
sample of other groups.

\begin{figure}
\psfig{figure=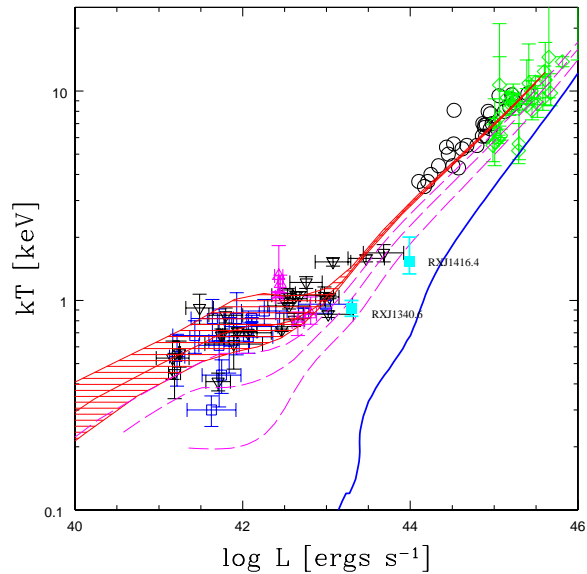,height=8.0cm,angle=0}
\caption{The X-ray bolometric luminosity-temperature relation for
groups and clusters at low redshifts, adapted from Babul \etal (2002).
The solid squares (labelled) are the two fossil groups with sufficient
$ROSAT$ counts to measure a temperature; both have excess luminosities
for their temperature (RXJ1340.6+4018 is described in Jones \etal
2000b).  The lines are predictions of the Babul \etal (2002)
analytical model. The heavy, solid curve is the prediction assuming no
preheating and isothermal gas at the virial temperature appropriate
for the mean redshift of formation of a given halo mass.  The hatched
region shows the prediction for a preheating entropy of
$kTn_e^{-2/3}\approx 427$ keV cm$^2$; the width of the region reflects
the range of formation redshifts. The three dashed curves show the
predictions for entropy levels of 300, 200 and 100 keV cm$^2$
respectively.  The two fossil groups are consistent with the 100 keV
cm$^2$ curve.  The data are from Markevitch (1998, circles), Allen \&
Fabian (1998, diamonds), Ponman \etal (1996, squares), Mulchaey \&
Zabludoff (1998, triangles) and Helsdon \& Ponman (2000, inverted
triangles) with luminosities corrected to the virial radius (see Babul
\etal 2002).  No error bars are given for the Markevitch (1998) data,
since they are smaller than the symbols.  All luminosities have been
corrected to H$_0$=75 km s$^{-1}$ Mpc$^{-1}$.}
\end{figure}

\subsection{The L$_X$-L$_{opt}$ relation}

In Fig 4 we plot the bolometric X-ray luminosities of the fossil
groups against the total optical luminosities of the central galaxies
(filled circles).  There is a strong correlation (correlation
coefficient 0.978, corresponding to a probability of no correlation of
$<$10$^{-7}$).  The best-fit relation for the fossils is
log$L_{X,42}$=(2.29$\pm$0.24)log$L_{R,11}$ + (0.55$\pm$0.14) + 2.58log\h50, where
$L_{X,42}$ is the bolometric X-ray luminosity in units of 10$^{42}$
erg s$^{-1}$, and $L_{R,11}$ is the R-band optical luminosity in units
of 10$^{11}$ L$\sol$.  The correlation implies a link between the
X-ray properties of the group and the central galaxy properties, and
is discussed further in Section 5.3.

A comparison with a similar relationship found for a sample of X-ray
bright groups by Helsdon \& Ponman (2003), and plotted as crosses in
Fig 4, is revealing. The fossil groups have much higher X-ray
luminosities than the Helsdon \& Ponman groups for their optical
luminosities, but a consistent slope; Helsdon \& Ponman find a slope
of 2.7$\pm$0.4.  A key difference between the datasets is that the
Helsdon \& Ponman points are for the {\it total}\/ optical luminosity
of the group, rather than of just the brightest galaxy.  A second
difference is that the $L_X$ values for the fossil groups have been
extrapolated beyond the detection radius, whereas the Helsdon \&
Ponman values have not.  The arrows on the plot indicate the likely
typical 
corrections for these differences.  Even assuming an overestimate of a
factor of 2 for the average increase in the X-ray luminosities of the
normal groups (Helsdon \& Ponman 2000), and increasing the fossil
group central galaxy optical luminosities by a factor of 1.5 to obtain
total group luminosities (as found by Jones \etal 2000b), the fossil
group X-ray luminosities are still significantly higher, by a factor of
$\approx$5.

\begin{figure}
\psfig{figure=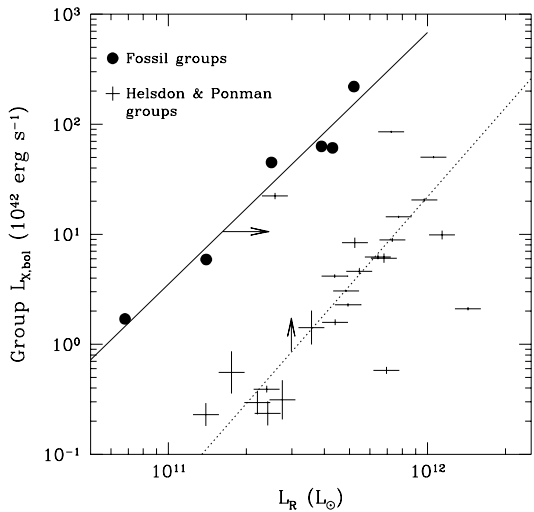,height=8.0cm,angle=0}
\caption{Group bolometric X-ray luminosity versus optical luminosity.
Filled circles show the fossil groups (5 from this paper plus one from
Jones \etal 2000b), for which $L_{opt}$ is that of the central
galaxy. The horizontal arrow shows the effect of correcting to the
group $L_{opt}$ by applying a factor of 1.5. The crosses show the
sample of X-ray bright groups of Helsdon \& Ponman (2003), for which
$L_{opt}$ is that of the group (corrected to $R$ assuming $B-R$=1.4).
The vertical arrow shows the approximate effect of extrapolating the
Helsdon \& Ponman X-ray luminosities to the virial radius. Best-fit
lines are also shown.  }
\end{figure}

\subsection{cD halos?}

We have investigated whether the luminous central galaxies have the
extended envelopes of cD galaxies.  A de Vaucouleurs r$^{1/4}$ profile
is a good fit to the surface-brightness profiles of all of the central
galaxies out to the largest radii at which they are detected
(corresponding to $\mu_R\approx26.5$ mag arcsec$^{-2}$).  No excess
above this profile at $\mu\lesssim$ 25 mag arcsec$^{-2}$,
characteristic of cD galaxies, is observed.

Residual images, after subtraction of the best fit elliptical models,
show no morphological signs of merger activity (eg. shells or tidal
tails), suggesting that the last major merger probably occurred
several Gyr ago.

\section{Discussion}

We have used well defined selection criteria to construct a sample of
fossil groups of galaxies based on a flux limited, complete survey of
extended X-ray sources. The presence of hot gas suggests that the
systems are gravitationally bound.  The constraints we have applied
are that $L_X\geq$10$^{42}$ \ergps (to avoid individual galaxies) and
\m12$\geq$2.0 mag, to ensure the presence of a dominant brightest galaxy.
What general properties do we observe?

Firstly, \m12 is not infinite; we have spectroscopically confirmed
group members at fainter magnitudes to show that these are indeed
systems of galaxies.  Secondly, all the brightest galaxies are giant
ellipticals and are located at, or consistent with, the
X-ray centroids.  Thirdly, the central galaxies are all optically
luminous, with total luminosities similar to the brightest galaxies in
clusters.  
NCG1275 in the Perseus
cluster has a relatively small cD envelope and has $M_R\approx
-24.5+5log(\h50)$, fainter than the most luminous fossil group galaxy
studied here, which has an absolute magnitude ($M_R=
-25.0+5log(\h50)$) similar to that of central cluster galaxies in rich
clusters (eg. Dressler 1978). 
In addition, the brightest galaxies in the poor AWM \& MKW
clusters have a similar range of optical luminosities as the central
fossil galaxies (Thuan \& Romanishin 1981).
A high luminosity of the brightest
galaxy was not a selection criterion; only a large luminosity gap was
required. Finally, we note that although (by definition) there are no
L$^*$ galaxies at r$<$0.5r$_{vir}$, at larger radii there are L$^*$
confirmed group members in at least some systems.

We interpret fossil groups as old, undisturbed systems in which the
L$^*$ galaxies within $r<0.5r_{vir}$ have merged to form the luminous
central elliptical galaxy observed today.  The age, derived from the
timescale for dynamical friction to cause L$^*$ galaxies to spiral in
toward the group centre, is $\sim$4 Gyr (Jones \etal 2000b).

\subsection{How numerous are fossil groups compared to BCGs?} 

Fossil systems represent 8\%-20\% of all systems of the same X-ray
luminosity (at $L_X>10^{43}$ \ergps and at $L_X>10^{42}$ \ergps),
based on a comparison with the integrated local XLF of Ebeling \etal
(1997). These fractions are also consistent with the cluster numbers
within the WARPS survey.  Based on a comparison with the integrated
XLF of HCGs of Ponman \etal (1996), fossil groups are $\approx$2-4
times more numerous than Hickson groups.

Could fossil groups be the site of formation of a large fraction of
BCGs, before infall of the fossil groups into clusters? Each fossil
system certainly contains a giant elliptical galaxy of an optical
luminosity similar to that of BCGs.  There are also easily enough
fossil systems. Fossil groups are at least as numerous as all typical
and rich clusters combined.  Even if only the most luminous fossil
systems of $L_X>10^{43}$ \ergps are considered, their space density is
the same as all clusters with $L_X>10^{44}$\ergps (0.5-2 keV)
ie. $\gtrsim$0.2$L_X^*$, and corresponding to $T_X\gtrsim$3 keV or
M$_{tot}\gtrsim 3\times10^{14}$ \Msol. Including low-luminosity fossil
systems (of $L_X>10^{42}$ \ergps) gives a space density $\sim$10 times
that of these clusters.

Groups were suggested as the site of galaxy merging which forms bright
cluster ellipticals by Aarseth \& Fall (1980). This process is also
assumed to occur in the semi-analytical galaxy formation models of
eg. Kauffmann, White \& Guiderdoni (1993).

Is the space density of the fossil groups consistent with them being
the origin of cluster BCGs?  We assume that both the fossil groups and
the BCGs have a common origin.  In keeping with our assertion that
fossil groups are old, undisturbed groups in which dynamical friction
has resulted in the merging of most of the L$^*$ galaxies, we
associate the present-day fossil groups/BGCs with 10$^{13}$-10$^{14}$
\Msol~ systems that ``formed'' at z$\gtrsim 1$ and which have essentially 
evolved relatively quietly  for most of the duration towards the present.
Noting that the notion  of ``formation time'' is ambiguous, we follow 
Balogh \etal (1999) and define it
as the epoch at which 70-75\% of the system's final mass 
is  assembled within a single halo.  This definition ensures that
post formation, the systems will typically evolve in a dynamically
quiet manner.  Numerical simulation  studies (see, for example, 
Navarro, Frenk \&  White 1995) indicate that once 70-75\% 
of a system's mass is assembled, the rest of the mass generally 
accretes via minor merger events that have no significant impact 
on structure of the halo.  Using the Extended Press-Schecter (EPS) 
formalism (Lacey \& Cole 1993), we find that bulk of the systems 
present at z$\approx$1 will have formed at 1$<$z$<$1.5 (ie. in the 
previous 1 Gyr).   Not all 
of these systems will evolve into fossil groups/BCGs.  We focus 
in on those systems that enjoy a quiet evolution from $z$=1-1.5 to 
$z$=0.3, that is for at least $\approx$4 Gyrs, so that dynamical
friction has enough time to act. 
The comoving number density of such systems is n(z=0.3)$\approx$
8x10$^{-6}$\h50$^3$Mpc$^{-3}$.  Of these, we identify
as fossil groups those that continue their quiet existence to the
epoch of observation.   At z=0.1, the comoving number density of 
such systems is n(z=0.1)$\approx$3x10$^{-6}$\h50$^3$Mpc$^{-3}$, 
which is comparable to our estimate of the number density of 
fossil groups (4x10$^{-6}$\h50$^3$Mpc$^{-3}$).  According to 
the EPS formalism, 8\% of the fossil groups present at z=0.3 are
destined to fall into clusters (of mass $>$10$^{14}$ \Msol) by z=0.
We identify these systems as BCGs and their corresponding 
predicted comoving number density is 6.4x10$^{-7}$\h50
$^3$ Mpc$^{-3}$.  The observed number density of BCGs can be
estimated from the cluster number density.  The current number 
density of clusters of mass $>$10$^{14}$\Msol~ (or $T_X>$2 keV) is
$\approx$1.6x10$^{-6}$\h50$^3$Mpc$^{-3}$ (Henry 2000), of which
$\approx$35-80\% have dominant central galaxies (depending on the
definition used, Jones \& Forman 1984).  Thus the density of 
BCGs is $\sim$(6-13)x10$^{-7}$\h50$^3$Mpc$^{-3}$, which is in 
good agreement with the prediction.  The concordance between
the various predicted and observed number densities is tantalizing
and we are preparing a much more detailed theoretical analyses of
the hypothesized association between fossil groups and BCGs using
both numerical simulations and the merger tree/analytic dynamical
evolution formalism of Taylor \& Babul (2001).

In this scenario, the cD halo often found around BCGs would be added after infall.
It could arise from stars stripped from galaxies within the cluster 
environment (eg. Lopez-Cruz \etal 1997). Speeding cDs  (eg. Zabludoff \etal 1993)
could be explained by fossil group infall.

Ghigna \etal (2000) have performed a high-resolution simulation of a
cluster of mass similar to that of the Virgo cluster within a cold
dark matter Universe, and thus including infall from its surroundings.
The core of the dark matter halo corresponding to the central galaxy
is largely assembled at high redshift (z=1-3) from the merger and
accretion of $\sim$12 halos with masses corresponding approximately to
L* galaxies (v$_{circ}\sim200-300$ km s$^{-1}$).  The mergers of these
halos occur before the formation of the cluster.

\subsection{Excess X-ray luminosities}

Possible explanations for the excess X-ray luminosities in Figs 3 \& 4
include:

(a) Point source contamination. The $ROSAT$ PSPC resolution of
$\approx$30\arcsec~ is insufficient to resolve all but the brightest
point sources. In the systems with the largest angular extent, the
contribution from AGN (either in the central galaxy, or in the
background), can be limited to $\approx$10-20\%. Thus for the two
systems which are measured to be overluminous for their temperatures
(RXJ1416.4+2315 and RXJ1340.6+4018), point source contamination is
unlikely to be responsible.  For some other fossil groups, such as
RXJ1331.5+1108, the point source contribution may be significant.

(b) Cooler central gas. If the groups are undisturbed and old, gas in the
innermost region may be cooling, increasing the X-ray luminosity and
lowering the mean temperature.  Individual systems can be offset from
the $L_X-T$ relation because of a large cool gas
contribution. Jones \etal (2000b) found a short cooling time of
$\sim$1Gyr at $r<6$ arcsec in RXJ1340.6+4018, and Vikhlinin \etal
(1999) found evidence for cool gas in 1159+5531. The cool gas
contributions to the total X-ray luminosities are
unknown, but by analogy with clusters, they are unlikely to account
for the factor $\approx$5 increase in $L_X$ seen in Fig 4.
 
(c) A low central entropy. The value of the gas entropy at
$r=0.1r_{vir}$ in low temperature groups has been found to have an
approximately constant value from group to group (the `entropy floor'
of Ponman \etal 1999 and Lloyd-Davies \etal 2000).  Here the entropy
is defined as $S=kT/n_e^{2/3}$. The `entropy floor' can be explained
by energy injection in addition to the compression and shock heating
caused by the dark matter gravitational potential (eg. Kaiser 1991,
Evrard \& Henry 1991).  The origin and epoch of the energy injection
is unknown, but may be related to supernova galaxy winds or AGN jets
(eg. Valageas \& Silk 1999, Bower \etal 2001).  A lower central
entropy than normal would allow the gas to achieve higher densities,
accounting for the high luminosity observed.  Fossil systems may thus
represent evidence for a distribution of central entropy values from
group to group as opposed to a constant value.

We illustrate this point in Fig 3 using the relatively simple analytic
model of Babul \etal (2002) to predict the $L_X-T$ relation for
various values of the entropy of the pre-heated gas. The model uses
the Lacey \& Cole (1994) mass distribution function to derive the
distribution of formation redshifts, assumes an isentropic core (the
size of which is a function of mass), outside of which accreting gas
is shock heated, and uses the Raymond \& Smith (1977) emissivities to
predict X-ray luminosities. The normalisation and shape of the
observed group and cluster $L_X-T$ relation at z=0 are well matched for a
pre-heated gas entropy of $\approx$427 keV cm$^2$ (hatched region in
Fig 3) for this particular model.  
The two fossil group points lie near lower values of the
pre-heated gas entropy (shown by the dashed lines), but not as low as
the prediction for no pre-heating (solid line), where the (isothermal)
temperature is given by the virial temperature at the epoch of
formation.

A lower central entropy could be due to either an early formation
epoch or simply to less energy injection.  If the energy injection in
some fossil groups occurred at high redshifts, earlier than in most
groups, and thus at higher densities, then a given injected energy
would produce a lower entropy in these systems (Ponman \etal 1999).
An early epoch of formation is consistent with the hypothesis that
fossil groups are generally old, undisturbed systems in which the
bright galaxies have merged. Alternatively, if the groups were formed
with only one very luminous, massive galaxy, then the high escape
velocity from that galaxy may have prevented supernova winds reaching
the intra-group medium and thus reduced the specific energy injection.

(d) The offset in Fig 4 between the fossil groups and the Helsdon \&
Ponman groups may alternatively be due to low (by a factor $\approx$2.5) 
optical luminosities of the fossil groups, rather than high X-ray
luminosities. This  would imply abnormally high mass-to-light ratios
and low star-formation efficiencies (Vikhlinin \etal 1999).  
Comparing optical luminosities with X-ray temperatures, the two
fossil groups with measured X-ray temperatures do indeed have low optical
luminosities for their temperatures when compared with the Helsdon \&
Ponman sample (a factor $\approx$1.5 below the mean relation), but
this is well within the scatter of the Helsdon \& Ponman data. More
detailed X-ray and optical observations will considerably clarify the
situation.

The origin of the excess X-ray luminosities will become clearer when
studies using Chandra data, in progress, are complete. 

\subsection{The origin of fossil groups}

The unusual distribution of galaxy light in fossil groups could be due
to merging of L* galaxies, as we have proposed (Ponman \etal 1994,
Jones \etal 2000b), or simply due to an unusual distribution of galaxy
masses when the groups formed. Although there is as yet no direct
evidence of merger activity in fossil groups, there are several
arguments supporting the merger origin.

In one sense, the luminous galaxies in fossil groups are extreme
versions of brightest cluster galaxies (BCGs), since they emit a very
high fraction of the total group light (70\% in RXJ1340.6+4018; Jones
\etal 2000b).  The optical luminosities of BCGs have been shown to be
not drawn from the luminosity function of cluster ellipticals (Sandage
1976; Tremaine \& Richstone 1977; Bernstein \& Bhavsar 2001),
suggesting a different origin for at least some of the BCG light.

To test the likelihood of \m12$>$2.0 appearing by chance, we sampled
a  Schechter function with the parameters of the 
composite luminosity function of MKW/AWM clusters (Yamagata \&
Maehara 1986).  We performed 10,000 Monte Carlo simulations with
absolute magnitudes selected at random from the Schechter function
distribution. The number of simulated luminosity functions with
\m12$>$2.0 was 0.4\%$\pm$0.06\%, significantly lower than the 8\%-20\%
of clusters and groups of comparable luminosity found to be fossil
systems.

There are two pieces of observational evidence that support the
merger origin. Firstly, the gap in the galaxy luminosity function 
at L$^*$ in
the central regions of the systems matches the predicted effect of
dynamical friction, since the most massive galaxies are predicted to
fall into the centre earliest. The combination of the lack of  L$^*$ 
galaxies and the very luminous central galaxy is 
very suggestive of merging. Formation of the systems with the 
luminosity distribution observed today would require  the formation of 
a single  luminous galaxy to be accompanied by a deficiency of less
luminous galaxies. 

Secondly, there is a strong correlation between the group X-ray
luminosity and the central galaxy optical luminosity, when what is
expected is a relation with the {\it total}\/ group optical
luminosity.  Indeed, the slope is consistent with that found for
normal groups when plotted against the total group optical
luminosity. This suggests that, unusually, the luminosity of the
brightest galaxy is strongly related to the global group properties.
This would arise naturally if the central galaxies were a result of
multiple merging.  Galaxy merging (for the BCGs in low L$_X$ clusters)
has also been suggested to explain the differences in evolution found
for BCGs in host clusters of low and high L$_X$ (Burke, Collins \&
Mann 2000, Brough \etal 2002, Nelson \etal 2002).

\subsection{An evolutionary sequence}

Fossil groups are more common than HCGs of the same $L_X$ by a factor
of $\approx$2-4.  In addition, from the discussion above we find that
groups in general are $\sim$8 times more numerous than fossil groups.
We assume a simple evolutionary sequence. If loose groups form at an
early epoch, with the L* galaxies within them evolving via HCGs into
fossil groups over a Hubble time, leaving the X-ray luminosity almost
unchanged, then the relative space densities of the three different
types of system give information on the transformation rates.
We use a very simple model in which the number of groups in any one 
state is proportional to the difference between the formation and 
destruction rates of that state.  The observed relative abundances are obtained
if the formation rate of HCGs is $\approx$1.3 times the formation rate
of fossil groups, and the formation rate of loose groups is $\approx$6
times the formation rate of HCGs. A relatively slow rate of formation
of fossil groups is consistent with the long dynamical friction
timescale of L* galaxies.

\subsection{The environments of luminous galaxies}

In the local Universe, the majority of very luminous galaxies are not
found in poor or rich cluster environments, but rather in binary
systems or groups (Cappi \etal 1998, Giuricin \etal 2001).  Moreover,
of the 113 very luminous galaxies with $M_B<$ -22.5 + 5log(\h50)
(ie. $L>4L^*$) in the Southern Sky Redshift Survey 2, Cappi \etal
(1998) estimate that $\approx$5-20\% are in groups dominated by the
single luminous galaxy.  The optical luminosities of these luminous
galaxies are similar to those of the central galaxies in our fossil
groups. Most of the very luminous galaxies of Cappi \etal are spirals. However, of
the very luminous early-type galaxies, $\approx$50\%-60\% are not in
rich or poor clusters.  Colbert, Mulchaey \& Zabludoff (2001) have
also found several isolated, luminous ($M_B<$ -22.5 + 5log(\h50))
early-type galaxies.

Although usually no redshift information on the fainter members of the
Cappi \etal groups is available, several of these systems appear to
have a distribution of optical galaxy luminosities very similar to
that of our fossil groups (see also Cappi \etal 2000).  In addition,
very luminous galaxies in general (whether in groups or not) have a
clustering amplitude higher than that of less luminous galaxies, and
similar to that of rich groups (Giuricin \etal 2001).  Because
clustering amplitude increases with system mass, from galaxies to
clusters, Cappi \etal (1998) suggested that very luminous galaxies may
be associated with dark halos of group mass.

However, not all very luminous elliptical galaxies are in fossil
systems with extended X-ray halos of high luminosity.  O'Sullivan,
Forbes \& Ponman (2001) have investigated L$_X$ as a function of
L$_{opt}$ and environment for nearby early-type galaxies.  They find a
few examples of optically very luminous early-type galaxies (which are
not at the centres of groups or clusters) with $L_X$ both above {\it
and}\/ below the limit of 10$^{42}$ \h50$^{-2}$ erg s$^{-1}$, the
limit we use here to define fossil groups.  Here we again define very
luminous as $M_B<-22.5$+5log(\h50). Thus examples of optically very
luminous galaxies with relatively low L$_X$ do exist. However, a
larger sample is required to accurately measure the fraction of
optically selected very luminous elliptical galaxies which are in
systems similar to the fossil groups described here.

The fraction of very luminous elliptical galaxies which were formed in
fossil groups is difficult to estimate.  The space density of all
galaxies more luminous than our faintest central galaxy
($M_R<-22.8$+5log(\h50), or $L_R>1.5L^*$) is $2\times 10^{-4}$
\h50$^3$Mpc$^{-3}$, based on the 2dF galaxy redshift survey luminosity
function of Norberg \etal (2002) and assuming $B-R$=1.5 mag.  This is
25 times the density of fossil groups, but only perhaps a quarter of
these may be early-type galaxies (Giuricin \etal 2001).
Evolutionary effects also complicate this comparison. For example, 
recent infall of L$^*$ galaxies into the cores of fossil groups would
remove them from the fossil category.
Our results do however suggest that at least a fraction of very
luminous elliptical galaxies formed via mergers in galaxy groups.

\section{Conclusions}

We have made an observational definition of a fossil system and
constructed an X-ray selected, flux-limited sample of 5 fossil groups
of galaxies with well defined selection criteria. The groups are
completely dominated by a central, luminous giant elliptical galaxy
with no cD halo.

Evidence for the luminous central galaxies being the result of
multiple mergers of L$^*$ galaxies includes: the gap in the galaxy
luminosity function at L$^*$, combined with the high luminosity of the
central galaxies and a low probability of obtaining
\m12$>$2 by chance,
and a strong correlation between the X-ray luminosity of the groups
and the optical luminosity of the central galaxies (with a slope
consistent with that found for $L_X$(group) versus $L_{opt}$(group)
for normal groups, although with an offset to higher X-ray
luminosities).

The X-ray luminosities of the fossil groups are well in excess of that
expected for their optical luminosities, and also for their X-ray
temperatures (for the two systems where $T_X$ can be measured). The
high X-ray luminosities may be caused by relatively cool gas in the
innermost regions or by a low
central gas entropy. A low central entropy could be the result of an
early epoch of formation, before most groups were formed, and
consistent with the interpretation of fossil systems as old,
undisturbed systems in which most L$^*$ galaxies have merged.

The fraction of all groups which are fossil systems of the same $L_X$
is 8\%-20\%.  Fossil groups are at least as numerous as all poor and
rich clusters combined (of $L_X>$10$^{43}$ erg s$^{-1}$) and thus are
potentially the site of formation of many brightest cluster galaxies,
before infall into clusters. However, only a fraction of all optically
luminous elliptical galaxies are found in X-ray bright group
environments.

\section{Acknowledgements}

We thank Steve Helsdon and Michael Balogh for discussions and
supplying data before publication, and the Isaac Newton Group staff on
La Palma for making INT service time observations. The anonymous referee
also made several useful suggestions. LRJ acknowledges
PPARC support during part of this work, AB acknowledges research
support from NSERC (Canada) and DJB acknowledges the support of NASA
contract NAS8-39073 (CXC).  We acknowledge use of the Digitized Sky
Survey, which was produced at the Space Telescope Science Institute,
the High Energy Astrophysics Science Archive Research Center
(HEASARC), provided by NASA's Goddard Space Flight Center, the
Leicester Database and Archive Service at the Department of Physics
and Astronomy, Leicester University, UK, and the NASA/IPAC
Extragalactic Database (NED), which is operated by the Jet Propulsion
Laboratory, California Institute of Technology, under contract with
NASA.

\section{REFERENCES}

\noindent
Albert C.E., White R.A., Morgan W.W., 1978, ApJ, 211, 309 (AWM)\\
Allen A.Q., 1973, Astrophysical Quantities. Athlone, London\\
Aarseth S.J., Fall S.M., 1980, ApJ, 236, 43\\
Avni Y., Bahcall J.N., 1980, ApJ, 235, 694\\
Babul A., Balogh M.L., Lewis G.F., Poole G.B., 2002, MNRAS, 330, 329\\
Balogh M.L., Babul A., Patton D.R., 1999, MNRAS, 307, 463\\
Bernstein J.P. \& Bhavsar S.P., 2001, MNRAS, 322, 625\\
Binney K., Tremaine S., 1987, Galactic Dynamics. Princeton Univ. Press,
 Princeton, NJ\\
Beers T.C., Kriessler J.R., Bird C.A., Huchra J.P. 1995, AJ, 109,
874\\
Borne K.D., Bushouse H., Lucas R.A., Colina L., 2000, ApJ, 529, L77\\
Bower R.G., Benson A.J., Lacey C.G., Baugh C.M., Cole S., Frenk C.S., 2001,
 MNRAS, 325, 497\\
Brough S., Collins C.A., Burke D.J.,  Mann R.G., Lynam P.D., 2002, MNRAS, 329, L53\\
Burke D.J., Collins C.A., Mann R.G., 2000, ApJ, 532, L105\\
Cappi A., Da Costa L.N., Benoist C., Maurogordato S., Pellegrini P.S.,
 1998, ApJ, 115, 200\\
Cappi A., Benoist C.,  Da Costa L.N., Maurogordato S., 2000, proceedings of 
 IAP 2000 meeting, Constructing the Universe with Clusters of Galaxies,
 F. Durret \& D. Gerbal eds.\\
Dressler A. 1978 ApJ 223, 765\\
Ebeling H. \& Wiedenmann G., 1993, Phys. Rev., 47, 704\\
Ebeling H., Jones L.R., Perlman E.S., Scharf C.A., Horner D.,
Wegner G., Malkan M., Mullis C.R., 2000,  ApJ, 534, 133.\\
Ebeling H., Jones L.R., Fairley B.W., Perlman E., Scharf C., 
Horner D., 2001, ApJ, 548, L23.\\
Evrard A.E., Henry J.P., 1991, ApJ, 383, 95\\
Evrard A.E., Metzler C.A., Navarro J.F, 1996, ApJ, 469, 494\\
Fairley B.W., Jones L.R., Scharf C.A., Ebeling H., Perlman E.S.,
 Horner D., Wegner G., Malkan M., 2000, 315, 669\\
Ghigna S., Moore B., Governato F., Lake G., Quinn T., Stadel J., 2000
 ApJ, 544, 616\\
Giuricin G., Samurovic S., Girardi M., Mezzetti M., Marinoni C., 2001,
 ApJ, 554, 857\\
Hagen H.J., Engels D., Reimers D., 1999, A\&AS, 134, 483\\
Helsdon S.F., Ponman, T.J., 2000, MNRAS, 315, 356\\
Helsdon S.F., Ponman, T.J., 2003, MNRAS, in press, astro-ph/0212046 \\
Henry J.P. 2000, ApJ, 534, 565\\
Hunsberger S.D., Charlton J.C., Zaritsky D., 1998, ApJ, 505, 536\\
Irwin M., Maddox S., McMahon R., 1994, Spectrum, 2, 14\\
Jones C., Forman W., 1984, ApJ, 276,38\\
Jones L.R., Scharf C.A., Ebeling H., Perlman E., Wegner G.,
 Malkan M., Horner, D., 1998, ApJ, 495, 100\\
Jones L.R., Ebeling H., Scharf C.A., Perlman E., Horner, D., Fairley B.,
 Wegner G., Malkan M. 2000a, in Durret F. \& Gerbal D., eds., Constructing the Universe
 with clusters of galaxies. Institute 
 d'Astrophysique de Paris, Paris\\
Jones L.R.,  Ponman T.J., Forbes D.A., 2000b, MNRAS, 312, 139\\
Kaiser N., 1991, ApJ, 383, 104\\
Kauffmann G., White S.D.M., Guiderdoni, B., 1993, MNRAS, 264, 201\\
Lacey C., Cole S., 1993, MNRAS, 262, 627\\
Lacey C., Cole S., 1994, MNRAS, 271, 676\\
Landolt A.U., 1992, AJ, 104, 1\\
Lloyd-Davies E.J., Ponman T.J., Cannon D.B., 2000, MNRAS, 315, 689\\
Lopez-Cruz O., Yee H.K.C., Brown J.P,. Jones C., Forman W., 1997,
ApJ, 475, L97\\
Mason K., et al 2000, MNRAS, 311, 456\\
Matsushita K., Makishima K., Ikebe Y.,
 Rokutanda E., Yamasaki N., Ohashi T., 1998, ApJ, 499, 13\\
Matsushita K., 2001, ApJ, 547, 693\\
Mewe R., Lemen J.R., van den Oord G.H.J., 1986, A\&A, 65, 511\\
Morgan W.W., Kayser S., White R.A., 1975, ApJ, 199, 545 (MKW)\\
Mulchaey J.S., Zabludoff, A.I., 1999, ApJ, 514, 133\\
Mushotzky R.F., Scharf, C.A. 1997 ApJ, 482, L13\\
Navarro J.F., Frenk C.S., White S.D.M., 1995, MNRAS, 275, 56\\ 
Nelson A.E., Gonzalez A.H., Zaritsky D., Dalcanton J.J., 2002, ApJ, 566, 103\\
Norberg P., et al, 2002, MNRAS submitted. \\
O'Sullivan E., Forbes D.A., Ponman T.J., 2001, MNRAS, 328, 461\\
Perlman E.S., Horner D., Jones L.R., Scharf C., Ebeling H., Wegner G., 
 Malkan M., 2002,  ApJS, in press.\\
Ponman T.J., Allan D.J., Jones L.R., Merrifield M., McHardy I.M.,
 Lehto H.J., Luppino G.A. 1994, Nature, 369, 462 \\
Ponman T.J., Bourner P.D.J., Ebeling H., Bohringer H., 1996, MNRAS, 283, 690\\
Ponman T.J., Cannon D.B., Navarro J.F., 1999, Nat, 397, 135\\
Price R., Burns J.O., Duric N., Newberry M.V., 1991, AJ, 102, 14\\
Raymond J.C., Smith B.W., 1977, ApJS, 35, 419\\
Romer A.K., Nichol R.C, Holden B.P., Ulmer M.P.,
 Pildis R.A., Merrelli A.J., Adami C., Burke D.J.,
 Collins  C.A., Metevier A.J., Kron R.G., Commons K., 2000, ApJS, 126, 209\\
Sandage A, 1976, ApJ, 205, 6\\
Scharf C.A., Jones L.R., Ebeling H., Perlman E., Malkan M.,
 Wegner G. 1997, ApJ, 477, 79\\
Taylor J.E., Babul A., 2001, ApJ, 559, 716\\
Thuan T.X., Romanishin W., 1981, ApJ, 248, 439\\
Tremaine S.D., Richstone D.O., 1977, ApJ, 212, 311\\
Tripp T.M., Lu L., Savage B.D., 1998, ApJ, 508, 200\\
Valageas P., Silk J., 1999, A\&A, 350, 725\\
Vikhlinin A., McNamara B.R., Forman W., Jones C., Quintana H., Hornstrup A.,
 1998, ApJ, 502, 558\\
Vikhlinin A., McNamara B.R., Hornstrup A., Quintana H., Forman W.,
 Jones C., Way C., 1999, ApJ, 520, L1\\
White D.A., Jones C., Forman W., 1997, MNRAS, 292, 419\\
White R.L., Becker R.H., Helfand D.J., Gregg M.D., 1997, ApJ 475, 479\\
Yamagata T., Maehara H. 1986, PASJ, 38, 661\\
Zabludoff A.I., Geller M.J., Huchra J.P., Vogeley M.S., 1993, AJ, 106, 1273\\ 

\end{document}